\documentclass[aps,12pt]{revtex4} 
\usepackage{graphicx,epsf}

\def\1s0{{^1\!S_0^{++}}}
\def\bp{{\bf p}}
\def\bn{{\bf n}}
\def\bk{{\bf k}}

\def\CK{{\cal K}}
\def\CM{{\cal M}}

\begin{document}

\title{Relativistic effects in proton-induced deuteron break-up 
at intermediate energies with forward emission of a fast proton pair}

\author{L.P. Kaptari}
\altaffiliation{On leave of absence from Bogoliubov Lab.\ Theor.\ Phys., JINR, Dubna, Russia;
also at Department of Physics, University of Perugia,
and INFN Sezione di Perugia, via A. Pascoli, I-06100, Italy}
\author{B. K\"ampfer}\email{kaempfer@fz-rossendorf.de}
\author{S.S. Semikh} 
\altaffiliation{Bogoliubov Lab. Theor. Phys., JINR, Dubna, Russia}
\affiliation{Forschungszentrum Rossendorf, PF 510119, 01314, Dresden, Germany}
\author{S.M. Dorkin$^\ddag$}
\affiliation{Nuclear Physics Institute, Moscow State University, Moscow, Russia}

\begin{abstract}
Recent data on the reaction $pd \to (pp) n$ with a fast forward
$pp$ pair with very small excitation energy is analyzed within a covariant
approach based on the Bethe-Salpeter formalism. It is demonstrated that
the minimum non-relativistic amplitude  
is completely masked by relativistic effects,
such as Lorentz boost and the negative-energy $P$ components
in the $^1S_0$ Bethe-Salpeter amplitude of the $pp$ pair. 
\end{abstract}

\maketitle 

The investigation of hadronic processes at high energies, such as
reactions of protons scattering off deuterons,
provides a refinement of information about strong interaction at short distances.
Nowadays, large research programs of experimental studies of processes 
with polarized particles are in progress.
Important are setups with deuteron targets or beams
\cite{kox0,preliminar,cosyproposal}, since
the deuteron serves as a unique source of information on the neutron
properties at high transferred momenta, the knowledge of which allows,
e.g. to check a number of QCD predictions and sum rules. Additionally,
the hadron-deuteron processes can be considered
as complementary tool in investigating short-distance phenomena and 
also as a source of information unavailable in electromagnetic reactions.
Of interest is the study of nucleon resonances, 
checking non-relativistic effective models,
meson-nucleon theory, $NN$ potentials etc.
In this line is the investigation of the deuteron break-up reaction
with a fast $pp$ pair at low excitation energy, proposed
in \cite{cosyproposal} and with first results reported in \cite{preprint_komarov}.

One motivation for the experiment \cite{preprint_komarov}
was the possibility to investigate the off-mass shell effects in $NN$ interactions.
As predicted in \cite{preprint_komarov,preprint_uzikov}, 
at a certain initial energy of the beam protons,
the cross section should exhibit a deep minimum, corresponding to the node of the
non-relativistic $^1S_0$ wave function of the two outgoing protons,
provided the non-relativistic
picture holds and the off-mass shell effects can be neglected.
The recent data \cite{preprint_komarov} exhibits, however,
a completely different behavior: the cross section is smoothly decreasing; there is no
sign of a pronounced minimum. Accounting for corrections beyond the 
one-nucleon exchange mechanism improve the agreement with data, however
a quantitative description have not achieved \cite{preprint_komarov}.

It is clear, that the non-relativistic treatment
of the process becomes inadequate because of the  high virtuality of
the  proton in the deuteron at the considered kinematics.
More realistic approaches which take into account
relativistic effects and the off-mass shellness of the interacting nucleons are desired.
The Bethe-Salpeter (BS) formalism can serve as an appropriate approach to the problem
because the off-mass shellness of the nucleons is an intrinsic feature of the
BS equation. Moreover, the solution of the BS equation, being manifestly
covariant, incorporate genuine relativistic effects (Lorenz boosts,
negative-energy components etc.), hardly accessible within the Schr\"odinger formalism.
In the present note we use the BS approach to
analyze the data \cite{preprint_komarov} on deuteron break-up with
the emission of a fast forward $pp$ pair \cite{foonote_1}.
We pay particular attention on relativistic effects in the wave function of
two nucleons in the continuum.
The model is based on our solution of the BS equation for
the deuteron with a
realistic one-boson exchange kernel \cite{solution}. The final
state interaction of  the two protons is
treated also within the BS formalism by
solving the BS equation for the $t$-matrix within the
one-iteration approximation \cite{ourfewbody,nashi}.
In doing so, a big deal of off-mass shell effects and relativistic
corrections are taken into account already within the spectator mechanism.

\begin{figure}[hb]
~\centering
\epsfxsize 2.7in
\epsfbox{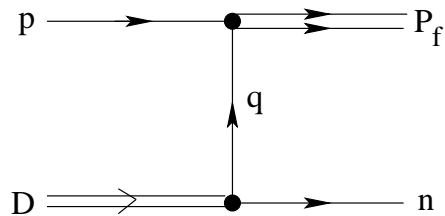}
~\vskip -6mm
\caption{Kinematics of the process (\protect\ref{reaction}).}
\label{pict1}
\end{figure}

Let us consider the process
\begin{equation} p\,+ d
\,=\,(p_1p_2)(0^0 ) +n(180^0)  
\label{reaction}
\end{equation}
at low excitation energy of the pair ($E_x \sim 0 - 3$ MeV)
and intermediate initial kinetic energies $T_p \sim 0.6 - 2.0$ GeV
corresponding to the conditions at the Cooler Synchrotron
COSY in the experiment \cite{cosyproposal,preprint_komarov}.
In the one-nucleon exchange approximation
this reaction can be represented by the diagram depicted in Fig.~\ref{pict1},
where the following notation is adopted: $p = (E_p,\bp)$ and
$n=(E_n,\bn)$ are the four-momenta of the incoming (beam) proton and outgoing
(not registered) neutron,
$P_f$ is the total four-momentum of the $pp$ pair, which is a sum of the corresponding
four-momenta of the detected protons, $p_1=(E_1,\bp_1)$, $p_2=(E_2,\bp_2)$.
The invariant mass of the $pp$-pair is
$M_{pp}^2=P_f^2=(2m+E_x)^2$, where $m$ stands for the nucleon
mass and  $E_x$ is the excitation energy.
Our calculations are performed in the laboratory
system where the deuteron is at rest.
For specific purposes, the center of mass
of the pair will be considered as well, where all relevant quantities are
superscripted  with asterisks.

A peculiarity of the processes (\ref{reaction})
is that the transferred three-momentum from the initial
proton to the second proton in the pair is rather high.
Moreover, from the kinematics one finds that
the momentum of the neutron is also high ($|\bn| \sim 0.3-0.5$ GeV/c),
which implies that, since the outgoing neutron is on-mass shell,
the proton inside the deuteron was essentially
off-mass shell before the interaction. Correspondingly,
it becomes clear that the process of $NN$ interaction in the upper part of the 
diagram is by far more complicate than the elastic interaction.
For instance, let us consider a typical kinematical situation, say
$|\bp |=1.22$ GeV/c, $\theta_1'\sim 4^0$, excitation energy $E_x = 3$ MeV
and $|\bp_1|=0.765$ GeV/c.
This means that the neutron three-momentum is $|\bn |\simeq 0.5$ GeV/c, i.e.,
the four-momentum of the off-mass shell proton was $q = (M_d-E_n,-\bn)$.
Now, if one supposes that in the upper
vertex there was an elastic process of two on-mass shell protons, then
only one kinematical quantity would be necessary to describe the process,
e.g., at given  $|\bp_1 |=0.765$ GeV/c the scattering angle would be $\sim 28^0$ in the
elastic kinematics (instead of $4^0$ in the full reaction);
or at  given scattering angle $4^0$,
the momentum of the elastically scattered proton would correspond to
$|\bp_1 |=1.21$ GeV/c ($|\bp_2| = 0.334$ GeV/c) instead of the detected
momentum  $|\bp_1|\simeq |\bp_2|\simeq 0.765$ GeV/c.
This demonstrates that the $NN$ interaction
in the upper part of the diagram has a quite complicate nature.
It can be considered as consisting at least of two steps: 
(i) an inelastic process which  puts the target nucleon on mass-shell, and
(ii) an elastic interaction in the $pp$ pair in the $^1S_0$ final state.
Since a large amount of the transferred energy is needed
to locate the second proton on mass-shell, 
the relativistic corrections can play a crucial role here.

The invariant cross section of the reaction (\ref{reaction}) reads
\begin{equation}
d^6\sigma\,=\,\frac{1}{(4\pi)^5\ \lambda(p,d)}\,
|M_{fi}|^2 \frac{d\bp_1\ d\bp_2}{E_1E_2}\, \delta\left( E_0+E_d-E_1-E_2-E_n\right),
\label{cross2}
\end{equation}
where $\lambda(p,d)$ is the flux factor, and $M_{fi}$ the invariant amplitude;
the statistical factor $1/2$ due to two identical particles (protons)
in the final state has been already included.
The covariant matrix element corresponding to the diagram in Fig.~\ref{pict1}
can be written in the form
\begin{equation}
M_{fi} = \bar u(s,\bn)(\hat n-m)\Psi_d(n) \left[
(\hat p_2+m) \overline{\Psi}_{^1S_0}(p) (\hat p_1-m) u(r,\bp) \right],
\label{ampl}
\end{equation}
where $u(r,\bp)$ ($\bar u(s,\bn)$) stands for
the Dirac spinor of the incident proton (outgoing neutron)
with the spin projection $r$ ($s$) and 3-momentum
$\bp$ ($\bn$),
$\Psi_{D(^1S_0)}$ denote the BS amplitudes for the deuteron and the $pp$-pair
in the continuum, respectively.
By using the spin angular basis to obtain the
partial decomposition of the deuteron BS amplitude
and the covariant form for the four partial components for the  $^1S_0$-state
\cite{nashi} of the $NN$ pair \cite{footnote_2},
the invariant amplitude $M_{fi}$ is
\begin{equation}
M_{fi}=(-1)^{\frac12 -r}\ \CK(^1S_0)\frac{1}{\sqrt{8\pi}(M_d-2E_n)}\left\{
\sqrt{2}C_{\frac12 s\frac12 -r}^{1\CM}
\left (G_S-\frac{G_D}{\sqrt{2}}\right )+
3\delta_{\CM,0}\delta_{s,r}\frac{G_D}{\sqrt{2}}
\right\},
\label{am1}
\end{equation}
where the contribution $\CK(^1S_0)$ from the upper part of the diagram in
Fig.~\ref{pict1} is
\begin{equation}
\CK(^1S_0) = \sqrt{\frac{E_p+m}{E_n+m}}
\left[
h_1\left( E_n+m-\frac{|\bn||\bp|}{E_p+m}\right)
-h_3\frac{M_d-2E_n}{m}\left(E_n+m+\frac{|\bn||\bp|}{E_p+m}\right)
\right].
\label{k1s0}
\end{equation}
$G_{S,D}$ denote the BS vertices for the deuteron,
$h_1,h_3$ are the non-vanishing  covariant partial components of
the $^1S_0$ configuration in the continuum. 
The relation of the amplitudes $h_i\ (i=1\ldots 4)$
to the partial solutions of the BS equation in the $NN$ center of mass
$^1S^{++}_0, ^1S^{--}_0, ^3P^{+-}_0, ^3P^{-+}_0$ can be found, 
e.g., in ref. \cite{nashi}. In what follows we neglect the contribution 
of the extremely small $^1S^{--}$ component, keeping
only the $^1S^{++}$ component as the main one and the $P$ components as 
the ones providing purely relativistic corrections.

It is easy to check that for the unpolarized particles the cross section factorizes
in two independent parts, i.e.,
$
\frac16
\sum\limits_{s,r,\CM} |M_{fi}|^2 \simeq \left| \CK(^1S_0)\right |^2\ \left(
u_S(\bn)^2 + u_D(\bn)^2
\right ),$
as it should be within the spectator mechanism with $^1S_0$ $(L_f=0)$
in the final state (see also discussion in ref.~\cite{ourdphi});
$u_S$ and $u_D$ are the BS deuteron $S$ and $D$ waves \cite{ourfewbody}

In our numerical calculations we use the deuteron BS amplitude from \cite{solution}.
For $^1S_0$ state one should solve the inhomogeneous BS equation 
in the $NN$ center of mass
to obtain the covariant amplitudes $h_{1,3}$. Note that in such a way
the effects of the Lorentz boost are  automatically taken into account
(see, e.g., \cite{ourfewbody}). The partial "$++$" components
of the BS vertices in the $NN$ center of mass,
for both the bound and the scattering states have a direct analogue
with the corresponding non-relativistic wave functions. Moreover, at low
intrinsic relative momenta the BS and non-relativistic wave functions
basically coincide. Hence, due to the low excitation energy
of the $pp$-pair, in expressing $h_{1,3}$ via the
partial amplitudes  at rest one may safely replace the "$++$"
vertex by its non-relativistic analogue. Relativistic effects are then
included in boosting the $^1S^{++}_0$ component to the deuteron center of mass
system and also by taking into account the contributions of $P$
components in $h_{1,3}$.
To find the $P$ waves we  solve the BS equation for the pair
in its center of mass within the one iteration approximation 
\cite{ourfewbody,ourcharge}, with the trial function as the exact
solution for the $t$ matrix within a separable potential \cite{plessas}.
By defining  new partial components  $g_i$ ($i=1\ldots 4$) as "connected amplitudes", 
i.e., the partial amplitudes without the free terms, we obtain for the $P$-waves
\begin{eqnarray}
g_3(k) = i\, g_{\pi NN}^2\,\,
\left[ \frac{M_{pp}}{\sqrt{\pi}}\,V_{31}(k,p^*)-
i\,\int\limits_0^\infty\,
\frac{dp\,p^2}{(2\pi)^3}\,V_{31}(k,p)\,\frac{g_1(p)}{M_{pp}-2\,E_p+i\epsilon}
\right],
\label{oia}
\end{eqnarray}
where the partial kernel $V_{31}$ is
$V_{31}(k,p^*)$
$=\displaystyle\frac{\pi m}{|\bp^*||\bk|E_{\bp^*}E_{\bk}}
\left\{
|\bp|^*Q_1(y)-|\bk|Q_0(y) \right\}$
with $Q_L(y)$ as Legendre function with the argument
$y=(|\bp^*|^2+|\bk|^2)+\mu_\pi^2-k_0^2)/(2|\bp^*||\bk|)$ 
\cite{ourcharge,nashi}.
In the first iteration the trial function $g_1(k)$
is expressed via the non-relativistic $t$ matrix \cite{plessas} as
\begin{eqnarray}
g_1(k)=i\, (4\pi)^{5/2}\,\frac{m}{2}\,\,t_{NR}(k,p^*),
\label{g1}
\end{eqnarray}
where the normalization of $t_{NR}$
corresponds to $
t_{NR}(p^*,p^*)=-\displaystyle\frac{2}{\pi m |\bp^*|}\, {\rm e}^{i\delta_0}\,\sin\,
\delta_0, $ where $\delta_0$ denotes the experimentally known phase-shift
of the elastic $pp$ scattering in the $^1S_0$ state. By using the 
Sokhotsky-Weierstrass formula for the Cauchy type integrals one finally obtains
\begin{eqnarray}
g_3(k) 
=\frac{i g_{\pi NN}^2}{\sqrt{\pi}}
\left[M_{pp} V_{31}(k,p^*) \left\{ 1 - \frac{i\pi m|\bp^*|}{2}
t_{NR}(p^*, p^*)\right\} 
+ m {\cal P} \int\limits_0^\infty 
dp p^2 V_{31}(k,p) \frac{t_{NR}(p, p^*)}{E_{p^*}-E_p}
\right], \nonumber
\label{oiafinal}
\end{eqnarray}
where, due to the separable stricture of the chosen $t_{NR}$, further calculations
of principal values of the relevant integrals can be carried out analytically.

\begin{figure}[hb]
~\centering
\epsfxsize 4.5in
\epsfbox{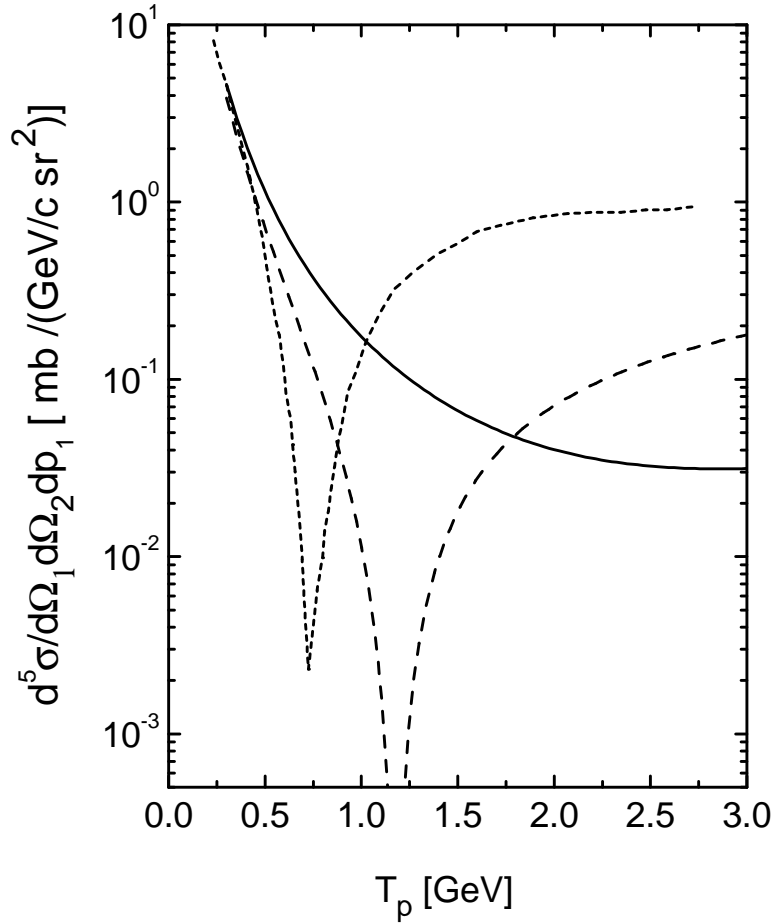}
\caption{Five-fold cross section as a function of the kinetic
energy $T_p$ of the incident proton. The dotted curve corresponds
to a non-relativistic calculation, i.e., to the case where only the
"$++$" components in the $^1S_0$ state are taken into account and any Lorentz
boost effects ignored. The dashed curve depicts results of calculations with
all relativistic effects in "$++$" components
taken into account. The solid curve is for the results of a complete calculation
with taking into account all the relativistic effects including
the contribution of $P$ waves in the wave function of the $pp$ pair.
The two protons are supposed to be detected in the strictly forward direction, i.e.,
$\theta_1=\theta_2 = 0^0$.}
\label{pict2}
\end{figure}

\begin{figure}[ht]
~\centering
\epsfxsize 4.5in
\epsfbox{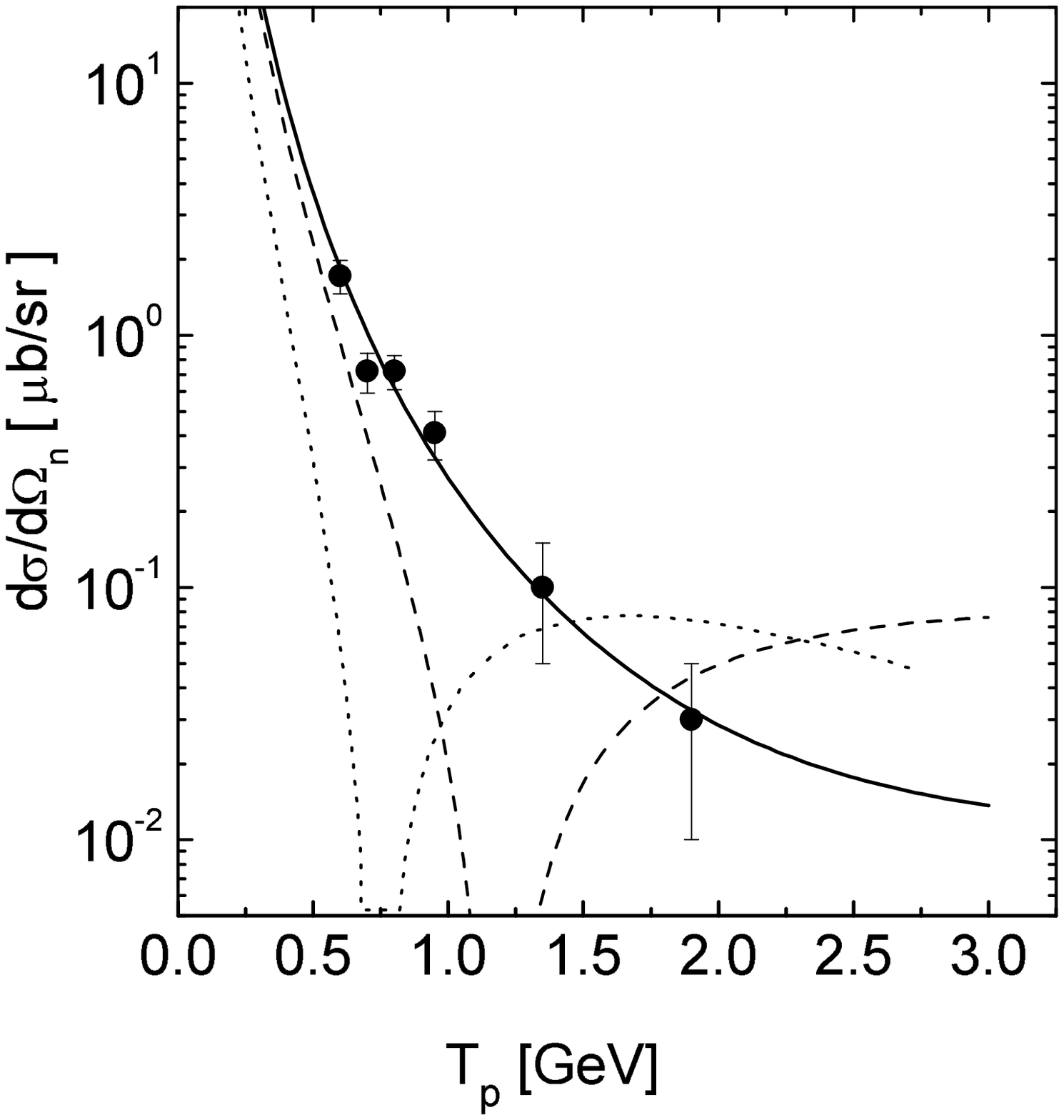}
\caption{Differential cross section in the center of mass
integrated over the excitation energy $E_x$
as a function of the  kinetic
energy $T_p$ of the incident proton.
Notation is as in Fig.~\protect\ref{pict2}. Experimental data
are from \protect\cite{preprint_komarov} }
\label{pict3}
\end{figure}

In Figs.~\ref{pict2} and \ref{pict3} we present results of numerical calculations of
the five-fold cross section ${d\sigma}/{d\Omega_1d\Omega_2d|\bp_1|}$
and the  two-fold cross section $ {d\sigma}/{d\Omega_{n}}$
(with $\Omega_{n}$ as the solid angle of the  momentum of the neutron in
the center of mass of the reaction),
integrated over the excitation energy in a range $E_x = 0 - 3$ MeV.
The calculations have been performed with our numerical solution for the
deuteron BS amplitude (inclusion of the $P$ components in the deuteron amplitude
leads to negligibly small corrections, therefore here we do not discus these
contributions).
The dotted curves are results of non-relativistic calculations, 
while the dashed curves include pure Lorentz boos effects, 
i.e., relativistic calculations with
including the $^1S^{++}_0$ component only. It is clearly seen that the
boost corrections are fairly visible: They cause a shift of the
minimum of the cross section. The agreement at low
initial energies becomes better, however, 
the cross section is still too small at large values of $T_p$.
Fig.~\ref{pict3} reveals that an account for only $"++"$ components
is not sufficient to describe data. An excellent description is achieved
by taking into account all the relativistic effects, including the contribution
of the negative-energy $P$ waves.
Note, that as demonstrated in refs.~\cite{ourfewbody,nashi}
in reactions of $pd$ and in threshold-near $ed$ disintegration,  
the inclusion of $P$ waves
exactly recovers the non-relativistic calculation with taking into account the
$N\bar N$ pair production effects. Hence, in our case this is a hint that covariant
calculations within the relativistic spectator mechanism contains already some
contributions beyond the one-nucleon exchange mechanism.

In summary we analyze the recent data \cite{preprint_komarov}
of the reactions $p d(p,n) \to (pp) n$ within a covariant approach
based on the Bethe-Salpeter formalism.
Relativistic effects (Lorentz boost, negative-energy $P$ components)
are important and responsible for the smooth decline of the cross
section, in contrast to predictions of non-relativistic models.

\acknowledgments
We are grateful to H.W. Barz for valuable discussions.
L.P.K. and S.S.S. acknowledge the warm hospitality of the nuclear
theory group in the Research Center Rossendorf.
The work has been supported by
Heisenberg - Landau program, BMBF grant 06DR921 and RFBR  00-15-96737.



\begin{thebibliography}{99}
\bibitem{kox0}
            S. Kox, E.J. Beise (spokespersons),
            TJNAF experiments 94-018 "Measurement of the Deuteron Polarization at
             Large Momentum Transvers in $D(e,e')D$ Scattering";
            http://www.jlab.org/ex\_prog;
            Phys. Rev. Lett. {\bf 84}, 5053, (2000);
            Nucl. Phys. {\bf A684} 521, (2001).
\bibitem{preliminar}
            E. Tomasi-Gustafsson, in Proc. of {\it XIV International Seminar
            On High Energy Physics Problems}, Preprint JINR No. E1,2-2000-166
            (Dubna, 2000).
\bibitem{cosyproposal}
           V.I. Komarov (spokesman) et al.,
         COSY proposal \#20 (updated 1999),
         ``Exclusive deuteron break-up study with
           polarized protons and deuterons at COSY'',
           http://ikpd15.ikp.kfa-juelich.de:8085/doc/Proposals.html.
\bibitem{preprint_komarov}
         V.I. Komarov  et al.,
         e-Print Archive: nucl-ex/0210017.
\bibitem{preprint_uzikov}
          Yu.N. Uzikov, V.I. Komarov, F. Rathmann, H. Seyfarth,
          e-Print Archive: nucl-th/0211001; \\
       Yu.N. Uzikov, J.Phys. {\bf G28}, B13 (2002).
\bibitem{foonote_1}
The considered reaction $p d \to (pp) n$
must not be confused with the charge exchange reaction $p d \to n (pp)$ 
with a slow pair in the final state.
\bibitem{solution}
         A.Yu. Umnikov, L.P. Kaptari, F.C. Khanna,
         Phys. Rev. {\bf C56}, 1700 (1997); \\
         A.Yu. Umnikov, L.P. Kaptari, K.Yu. Kazakov, F.C. Khanna, 
         Phys. Lett. {\bf B334}, 163 (1994);\\
         A.Yu. Umnikov, Z. Phys. {\bf A357}, 333 (1997).
\bibitem{ourfewbody}
        L.P. Kaptari, B. K\"ampfer, S.M. Dorkin, S.S. Semikh,
        Few-Body Systems {\bf 27}, 189 (1999);
        Phys. Rev. {\bf C57}, 1097 (1998);
        Phys. Lett. {\bf B404}, 8 (1997); \\
        P. Kaptari, B. K\"ampfer, A.Yu. Umnikov, F.C. Khanna,
        Phys. Lett. {\bf B351}. 400 (1995).
\bibitem{nashi}
        S.G. Bondarenko, V.V. Burov, M. Beyer, S.M. Dorkin,
        Phys. Rev. {\bf C58}, 3143 (1998);  e-Print Archive: nucl-th/9612047.
\bibitem{footnote_2}
Actually, when one particle
is on mass shell (the incident proton in our case), only two
partial amplitudes differ from zero corresponding to
positive values of the $\rho$-spin of the on mass shell particle.
\bibitem{ourdphi}
        L.P. Kaptari, B. K\"ampfer, S.S. Semikh,
        e-Print Archive: nucl-th/0212066.
\bibitem{ourcharge}
        L.P. Kaptari, B. K\"ampfer, S.S. Semikh, S.M. Dorkin,
        arXiv:nucl-th/021107;\\
        L.P. Kaptari, B. K\"ampfer, S.S. Semikh, S.M. Dorkin,
        Phys. Atom. Nucl.{\bf 65}, 442 (2002).
\bibitem{plessas}
           J. Haidenbauer, Y. Koike, W. Plessas,
          Phys. Rev. {\bf C33}, 439 (1986).
\end{thebibliography}
\end{document}